# Magnetoelectric control of topological phases in graphene

Hiroyuki Takenaka, Shane Sandhoefner, Alexey A. Kovalev, and Evgeny Y. Tsymbal*

*Department of Physics and Astronomy & Nebraska Center for Materials and Nanoscience,
University of Nebraska, Lincoln, Nebraska 68588-0299, USA*

Topological antiferromagnetic (AFM) spintronics is an emerging field of research, which involves the topological electronic states coupled to the AFM order parameter known as the Néel vector. The control of these states is envisioned through manipulation of the Néel vector by spin-orbit torques driven by electric currents. Here we propose a different approach favorable for low-power AFM spintronics, where the control of the topological states in a two-dimensional material, such as graphene, is performed via the proximity effect by the voltage induced switching of the Néel vector in an adjacent magnetoelectric AFM insulator, such as chromia. Mediated by the symmetry protected boundary magnetization and the induced Rashba-type spin-orbit coupling at the interface between graphene and chromia, the emergent topological phases in graphene can be controlled by the Néel vector. Using density functional theory and tight-binding Hamiltonian approaches, we model a graphene/$Cr_2O_3$ (0001) interface and demonstrate non-trivial band gap openings in the graphene Dirac bands asymmetric between the $K$ and $K'$ valleys. This gives rise to an unconventional quantum anomalous Hall effect (QAHE) with a quantized value of $2e^2/h$ and an additional step-like feature at a value close to $e^2/2h$, and the emergence of the spin-polarized valley Hall effect (VHE). Furthermore, depending on the Néel vector orientation, we predict the appearance and transformation of different topological phases in graphene across the 180° AFM domain wall, involving the QAHE, the valley-polarized QAHE and the quantum VHE (QVHE), and the emergence of the chiral edge state along the domain wall. These topological properties are controlled by voltage through magnetoelectric switching of the AFM insulator with no need for spin-orbit torques.

## I. INTRODUCTION

In the past decades, spintronics has been considered as a promising avenue to establish new frontiers in information technology by exploiting the spin degree of freedom [1]. Driven by this technological challenge, exploration of new spintronic phenomena has become one of the most active research topics in condensed matter physics. Recently, antiferromagnetic (AFM) spintronics has emerged as a subfield of spintronics, where the AFM order parameter known as the Néel vector was employed as the non-volatile state variable [2-5]. Due to being robust against magnetic perturbations and exhibiting ultrafast dynamics, antiferromagnets can serve as promising functional materials for spintronic applications.

In parallel with these developments, there has been increasing interest in materials and structures where quantum effects are responsible for novel physical properties, revealing the important roles of symmetry, topology, and dimensionality [6]. Among such quantum materials are graphene [7], topological insulators [8], Dirac and Weyl semimetals [9], and beyond [10]. The unique spin-dependent electronic properties of these materials are envisioned to open new perspectives for spintronic applications [11-13]. Among them, topological AFM spintronics is especially interesting, involving the interplay between the topological electronic states and antiferromagnetism [14-17].

The key ingredient of AFM spintronics, including its topological variant, is a possibility to control the Néel vector by external stimulus. Achieving this functionality in antiferromagnets is not straightforward as in ferromagnets where the control of the magnetic order parameter, i.e. magnetization, can be realized by an applied magnetic field, spin transfer torque or spin-orbit torque. Recently, it has been predicted that in antiferromagnets with two spin sublattices forming inversion partners, an electrical current can induce a non-equilibrium magnetic field, which sign alternates between the spin sublattices [18]. Such a staggered magnetic field generates an alternating sign in spin-orbit torque, which can trigger Néel vector switching. This prediction has been realized experimentally for CuMnAs [19] and Mn2Au [20] antiferromagnets, thus demonstrating the viable route for AFM spintronics.

While this progress is impressive, it is recognized that very large electric currents are needed to generate the required spin-orbit torques (>$10^6$ Acm$^{-2}$). Such currents would inevitably produce significant energy dissipation unfavorable for the desired low-power spintronics. From this perspective, it is beneficial to have the ability of controlling the Néel vector purely by electric fields through an applied voltage. Such a control has been demonstrated using magnetoelectric antiferromagnet $Cr_2O_3$ in the corundum structure (chromia) [21]. Due to the magnetoelectric nature of this material, by applying an electric field (in the presence of a finite magnetic bias), one can switch the Néel vector between two non-volatile states. Utilizing this functionality in topological AFM spintronics would provide both ultra-low power performance and ultrafast switching dynamics.

To this end, the voltage-switchable Néel vector in chromia could be exploited to manipulate quantum and topological properties of two-dimensional materials, such as graphene, via the proximity effect. The exchange coupling between chromia and graphene across the interface in a graphene/$Cr_2O_3$ hybrid structure would be mediated by the boundary magnetization, which is the intrinsic property of all magnetoelectric antiferromagnets [22-24]. The boundary magnetization is firmly coupled to the bulk AFM order so that switching of the Néel vector leads to its reversal. Due to being insensitive to the interface roughness, the boundary magnetization can serve as a robust voltage-controlled parameter to operate topological properties of graphene. Experimental efforts along these lines have indicated the potential of the graphene/$Cr_2O_3$ hybrid structure for realizing a magnetoelectric transistor [25].

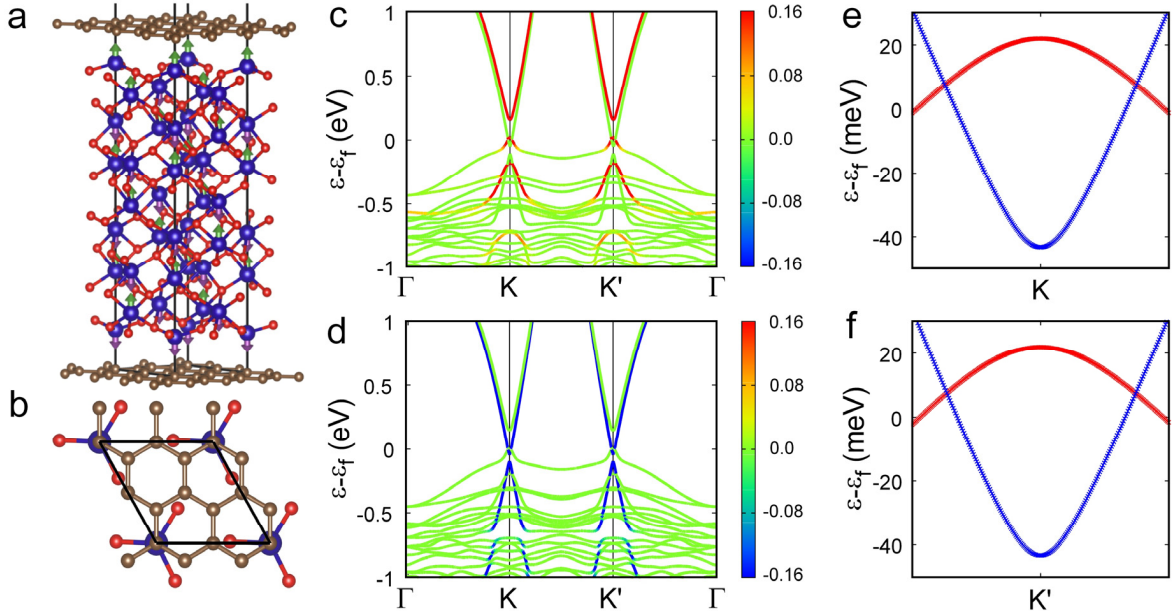

**Figure 1.** Atomic and electronic structure of the graphene/$Cr_2O_3$ (0001) interface without SOC. **a** The optimized atomic structure: side view. Blue, red, and gold balls indicate Cr, O, and C atoms, respectively. Color arrows denote up (green) and down (purple) spins in AFM chromia. **b** Top view of the atomic structure (**a**) showing graphene and surface Cr and subsurface O monolayers. **c, d** Electronic band structures projected to the top graphene layer for spin-up (**c**) and spin-down (**d**) electrons. The color contrast reflects the strength of the carbon $p_z$ orbital contribution weighted with the $s_z$ spin contribution in arbitrary units. **e-f** The spin-resolved bands originating from the graphene Dirac bands zoomed in near $K$ (**e**) and $K'$ (**f**) points. The same color coding is used as in **c** and **d**.

The appearance of topological effects in graphene, such as the quantum anomalous Hall (QAHE) [26] and the quantum spin Hall effect (QSHE) [27] requires spin-orbit coupling (SOC). It is known, however, that the intrinsic SOC in pristine graphene is extremely weak [28,29]. Yet, a sizable SOC in graphene can be induced by the proximity effect at the interface between graphene and other materials such as transition metal dichalcogenides [30,31]. In addition to SOC, the proximity effect can also induce a sizable spin-polarization in graphene when it is deposited on the surface of a magnetic insulator [32, 33], which can lead to the QAHE [34].

All these observations indicate that the proximity effect at the $Cr_2O_3$/graphene interface could produce SOC necessary for graphene to exhibit topological properties, which could be controlled by voltage though the boundary magnetization of chromia. Motivated by this idea, we use density functional theory (DFT) and model tight-binding approaches to explore spin- and orbital-dependent electronic and transport properties of the graphene/$Cr_2O_3$ (0001) interface. We show that the presence of a sizable SOC and exchange splitting in graphene induced by the proximity of chromia leads to the QAHE which changes sign with reversal of the $Cr_2O_3$ Néel vector. The broken symmetry between the A and B sublattices in graphene at the $Cr_2O_3$/graphene interface produces asymmetry between the $K$ and $K'$ valleys, resulting an unconventional QAHE and the emergence of the spin-polarized valley Hall effect. We predict the appearance and change of different topological phases in graphene across the 180° AFM domain wall and the emergence of the chiral edge state along the domain wall.

## II. RESULTS

### A. DFT calculations

We perform DFT calculations of the graphene/$Cr_2O_3$ (0001) interface, as described in Methods. A 2×2 unit cell of graphene excellently matches to the 1×1 unit cell of the $Cr_2O_3$ (0001) surface with a lattice mismatch of just 0.83% (see in Appendix A for computational details). By performing structural optimization (Supplementary Material [35]), we find the most energetically favorable atomic structure, which is shown in Figures 1 a,b. In this interface structure, one C atom lies atop the surface Cr atom, the distance between the two being 2.63 Å. Such a large distance indicates the weak bonding between graphene and the substrate consistent with the previous DFT calculations [32,34].

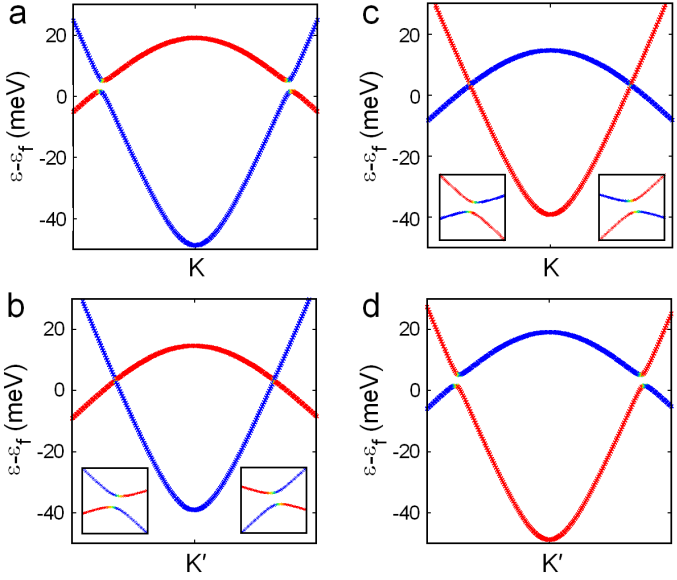

**Figure 2.** The DFT-calculated band structures of the graphene/Cr$_2$O$_3$ (0001) interface in the presence of SOC. **a,b** The band structures around the $K$ (**a**) and $K'$ (**b**) points. **c,d** Same as **a** and **b**, respectively, but for the reversed Néel vector in chromia. Color contrast reflects the $s_z$ spin contribution in the same way as in Figure 1. The insets in **b** and **c** are the zoomed bands to reveal small band openings of 0.7 meV.

The magnetic structure of bulk Cr$_2$O$_3$ represents a collinear antiferromagnetic configuration with Cr magnetic moments pointing along the (0001) direction (Figure 1a). The top surface Cr monolayer has parallel-aligned magnetic moments of $m_{Cr}$ = 2.8$\mu_B$, representing the boundary (surface) magnetization [23]. Reversal of the Néel vector in bulk chromia (which can be achieved by voltage) leads to the reversal of this surface magnetization.

We find that there is a sizable exchange splitting of the spin bands in graphene induced by the proximity of the surface magnetization of chromia. Figures 1 c,d show the band structure of the graphene/Cr$_2$O$_3$ (0001) interface calculated without SOC in vicinity of the Dirac point of pristine graphene. Here the bands are projected to the $p_z$ orbitals of the top graphene layer and their up- (Figure 1c) and down- (Figure 1d) spin weights are shown in color. It is seen that the spin bands are split by the induced exchange interaction. It is also seen that there is a splitting between bands of the same spin which is due to a staggered sublattice potential discussed below. Figures 2e and 2f zoom in on the spin-split bands originating from the graphene Dirac bands near the $K$ and $K'$ points, respectively. These figures reveal that the exchange splitting of the spin bands is about 60 meV, the band structures at the $K$ and $K'$ points are identical, and there are no gap openings at the band crossing points in the absence of SOC.

The broken inversion symmetry at the graphene/Cr$_2$O$_3$ interface gives rise to the Rashba-like SOC. The SOC mixes the up- and down-spin states and opens the gaps at the crossing points, as is evident from Figures 2a and 2b. We find that the band opening is about 3 meV near the $K$ valley (Fig. 2a) and is about 0.7 meV near the $K'$ valley (Fig. 2b). From comparison of Figures 2a and 2b, it is also notable that SOC reduces the effective spin splitting at the $K'$ valley. This difference in the band structure at the $K$ and $K'$ points is due to the variable bond length between Cr surface atoms and C atoms in the graphene $A$ and $B$ sublattices (Fig. 1b), resulting in the staggered potential, the staggered exchange interaction and the staggered SOC, as described by our tight-binding model below.

Switching the Néel vector in chromia is equivalent to the time reversal symmetry transformation. It is therefore expected that reversal of all the Cr magnetic moments in Cr$_2$O$_3$ would lead to swapping the bands structures between the $K$ and $K'$ points with simultaneous reversal of the spin character of the bands. This is exactly what we find by performing self-consistent DFT calculations in the presence of SOC for the graphene/Cr$_2$O$_3$ interface. From comparison of Figures 2 c,d and Figures 2a,b, we see that switching the Néel vector in chromia transforms the band structures between the $K$ and $K'$ points and at the same time reverses the spin character.

### B. Tight-Binding Model

To provide more insight into the proximity effect on the electronic band structure of graphene and to analyze its quantum transport behavior, we build a model tight-binding Hamiltonian as follows [31,36-39]

$$H = -t\sum_{\langle ij\rangle\alpha} c^\dagger_{Ai\alpha} c_{Bj\alpha} + it_{so}\sum_{\langle ij\rangle\alpha\beta} \hat{z}\cdot(\vec{\sigma}\times\hat{d}_{ij}) c^\dagger_{Ai\alpha} c_{Bj\beta} - \lambda_{nl}\sum_{\langle ij\rangle\alpha\beta}(\hat{m}\cdot\vec{\sigma}) c^\dagger_{Ai\alpha} c_{Bj\beta} + U\sum_{\mu=A,B}\eta_\mu \sum_{i\alpha} c^\dagger_{\mu i\alpha} c_{\mu i\alpha} + \sum_{\mu=A,B} i\lambda_{so,\mu}\sum_{\langle\langle ij\rangle\rangle\alpha} \nu_{ij} c^\dagger_{\mu i\alpha}\sigma_z c_{\mu j\alpha} - J_A\sum_{i\alpha}(\hat{m}\cdot\vec{\sigma}) c^\dagger_{Ai\alpha} c_{Ai\alpha} + J_B\sum_{i\alpha}(\hat{m}\cdot\vec{\sigma}) c^\dagger_{Bi\alpha} c_{Bi\alpha} + X\sum_{\mu=A,B}\sum_{i\alpha} c^\dagger_{\mu i\alpha} c_{\mu i\alpha}, \quad (1)$$

where $c^\dagger_{Ai\alpha}$ and $c_{Ai\alpha}$ are the creation and annihilation operators for site $i$ on sublattice $A$ with spin $\alpha$. The first term is the conventional tight-binding Hamiltonian for pristine graphene with hopping parameter $t$. The second term is the Rashba SOC, which includes parameter $t_{so}$, Pauli matrix vector $\vec{\sigma} = (\sigma_x, \sigma_y, \sigma_z)$, and the unit vector $\hat{d}_{ij}$ from site $j$ to $i$. The third term represents the non-local exchange interaction involving the spin-dependent hopping with $\hat{m} = (\sin\theta\sin\varphi, \sin\theta\cos\varphi, \cos\theta)$ being the unit vector of the surface magnetization. This term affects the slope of band curvatures by strength $\lambda_{nl}$. For simplicity, we assume $\varphi = 0$, but note that varying $\varphi$ only causes small changes in the band structure and does not affect our main results. The remaining terms describe the site-dependent interactions. The fourth term describes the staggered sublattice potential ($\eta_A = 1$ and $\eta_B = -1$) of strength $U$. The staggered potential opens the band gaps and rounds the bands near the Dirac points, but does not introduce differences in the band structure near the $K$ and $K'$ valleys. The fifth term describes the staggered SOC of amplitude $\lambda_{so}$, which involves the second-nearest neighbor hopping and has $v_{ij} = 1\ (-1)$ for clockwise (counterclockwise) hopping from cite $j$ to $i$. The band openings in question are strongly influenced by the staggered SOC. The next two terms correspond to the exchange interactions of strengths $J_A$ and $J_B$, which are assumed to be different on the A and B sublattices. The last term describes an overall energy shift. The summation over $\langle ij \rangle$ and $\langle\langle ij \rangle\rangle$ in Eq. (1) runs over all the nearest (next-nearest) neighbor sites, respectively. We note that although the model Hamiltonian contains many terms, all the terms appear to be necessary to quantitatively reproduce the DFT calculated band structures (Supplementary Figure S3). Details of the fitting procedure and the resulting fitting parameters are given in the Supplementary Material [35].

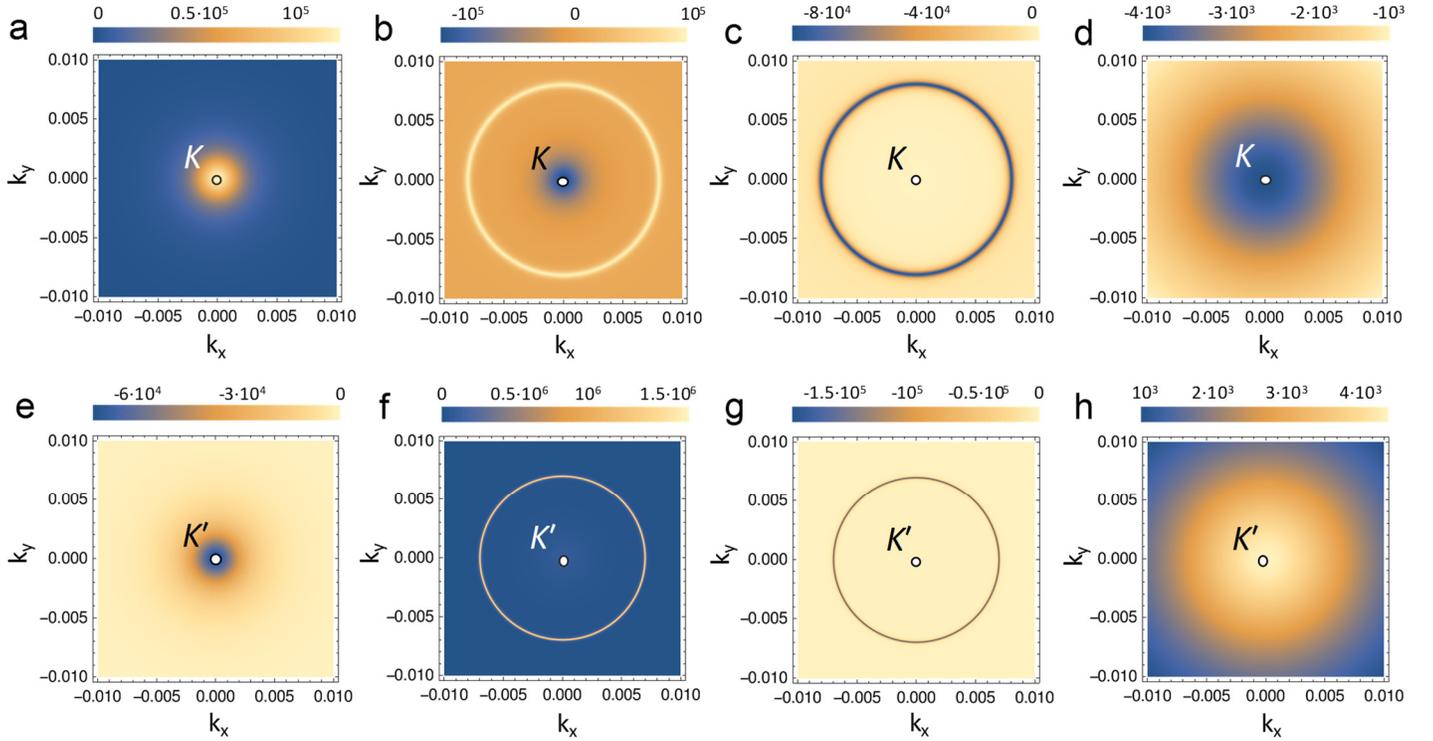

**Figure 3.** Color contour plots of the Berry curvature projected on the $k_x$-$k_y$ plane. **a-d** The Berry curvature for the four bands shown in Supplementary Figure S3a around the $K$ point. **e-h** The Berry curvature for the four bands shown in Supplementary Figure S3b around the $K'$ point. The bands are ordered from low to high energy. The origin of $x$- and $y$-axis is at the $K$ point in **a-d** and at the $K'$ point in **e-h**. Color bars quantify the Berry curvature.

### C. Berry curvature

Next, using Hamiltonian (1) with the fitted parameters, we analyze the QAHE, which is determined by the Berry curvature [9]

$$\Omega^n(\vec{k}) = -\sum_{n' \neq n} \frac{2\mathrm{Im}\langle \Psi_{n\vec{k}}|v_x|\Psi_{n'\vec{k}}\rangle\langle\Psi_{n'\vec{k}}|v_y|\Psi_{n\vec{k}}\rangle}{(\varepsilon_{n'\vec{k}} - \varepsilon_{n\vec{k}})^2}. \qquad (2)$$

Here $n$ is band index, $v_{x,y} = \partial H/\partial k_{x,y}$ is the velocity operator, and $\varepsilon_{nk}$ and $\Psi_{nk}$ are eigenvalues and eigenfunctions of the Hamiltonian at a given $k$-point within the Brillouin zone. For simplicity, we omit $x$ and $y$ indices in the notation for $\Omega^n(\vec{k})$ as well as for the anomalous Hall conductance (AHC), $\sigma$, below. In Figure 3, we show the calculated Berry curvature around the $K$ (Figs. 3 a-d) and $K'$ (Figs. 3 e-h) valleys for each of the four bands displayed in Supplementary Figures 3a and 3b, respectively. It is seen that the Berry curvature becomes very large on the circles around the $K$ and $K'$ points. The radii of these circles match the $k$ values, at which the band openings appear due to SOC (Figs. 2 a,b). The enhancement of $\Omega^n(\vec{k})$ at the $K$ and $K'$ points, especially pronounced for the lowest (Figs. 3 a,e) and highest (Figs. 3 d,h) energy bands, reflects their extrema at these points (Supplementary Figs. S3a and S3b [35]) resulting from the band opening produced by the staggered sublattice potential $U$.

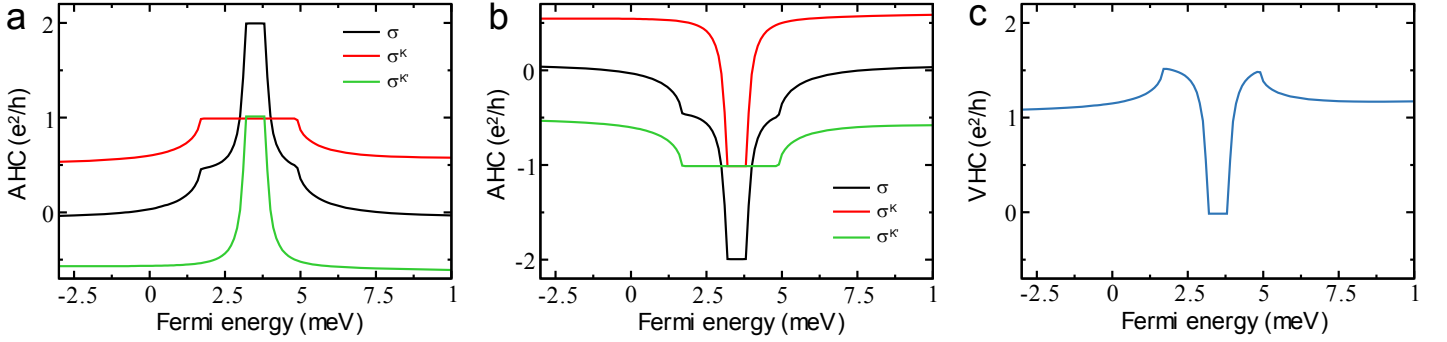

**Figure 4.** The calculated anomalous Hall conductance (AHC) in graphene as a function of the Fermi energy $\varepsilon_f$ for the Néel vector in chromia pointing up (**a**) and down (**b**). $\sigma$ is the total AHC, and $\sigma^K$ and $\sigma^{K'}$ are the partial contributions arising from the $K$ and $K'$ valleys, respectively. **c** Valley Hall conductance (VHC), $\sigma_v = \sigma^K - \sigma^{K'}$, as a function of $\varepsilon_f$. The results are obtained using Hamiltonian (1) with the parameters fitted to the DFT bands.

### D. Anomalous and valley Hall conductance

The AHC is determined by the Berry curvature, as follows[9]:

$$\sigma = \frac{e^2}{h}\frac{1}{2\pi}\sum_n \int_{BZ} f_n \Omega^n(\vec{k}) d^2k, \qquad (3)$$

where $e$ is elementary charge, $h$ is Planck's constant, and $f_n$ is the Fermi-Dirac distribution function. Figure 4a (black line) shows the calculated AHC, $\sigma$, as a function of the Fermi energy, $\varepsilon_f$. As expected, the AHC acquires the value of $\sigma = 2e^2/h$ when $\varepsilon_f$ lies within the energy region where there is a global energy gap in the system (i.e., where the band openings at the $K$ and $K'$ points overlap). When $\varepsilon_f$ lies far from this gap, the AHC tends to zero due the cancellation of the contributions from the $K$ and $K'$ valleys. It is notable that in the vicinity of $\varepsilon_f = 2$ meV and $\varepsilon_f = 5$ meV, i.e. above or below the smaller bandgap, the AHC exhibits an unconventional two-step-like feature associated with quantized conductance at $e^2/h$ for one valley but not for the other. This feature in the AHC appears due to different band gaps at the $K$ and $K'$ valleys. This is evident from Figure 4a, where we plot partial contributions to $\sigma$ arising from the $K$ and $K'$ valleys, $\sigma^K$ and $\sigma^{K'}$ (red and green lines in Fig. 4a), by integrating the Berry curvature over the respective $k$-space regions. Each $\sigma^K$ and $\sigma^{K'}$ has an exact quantized value of $e^2/h$, when $\varepsilon_f$ falls into the associated band gap. However, when $\varepsilon_f$ lies within the wider band of the $K$ valley, but above or below the band gap of the $K'$ valley, the contribution from the latter is $\sigma^K = e^2/h$, while the contribution from the former, $\sigma^{K'}$, drops down to about $-e^2/2h$ so that within this energy widow $\sigma$ appears to be close to a value of $e^2/2h$. The $\sigma^{K'}$ value of $-e^2/2h$ results from the integration of the Berry curvature of the lowest energy band around the $K'$ valley (Fig. 3e).

Thus, the AHC exhibits the unconventional two-step-like behavior due to the different band openings at the $K$ and $K'$ valleys.

Switching the Néel vector in chromia is equivalent to the time reversal symmetry operation. Since $\Omega^n(\vec{k})$ is odd with respect to time reversal symmetry, i.e. $\Omega^n(-\vec{k}, -\vec{s}) = -\Omega^n(\vec{k}, \vec{s})$, where $\vec{s}$ is the spin, it is expected that with reversal of the Néel vector $\vec{L}$, $\sigma$ will change sign and the partial contributions will transform as $\sigma^K(-\vec{L}) = -\sigma^{K'}(\vec{L})$ and $\sigma^{K'}(-\vec{L}) = -\sigma^K(\vec{L})$. This is exactly what we find from the calculation shown on Figure 4b, where reversal of the Néel vector in the Hamiltonian (1) was modelled by changing the angle of the surface magnetization from $\theta = 0$ to $\theta = 180º$. Since the Néel vector and the surface magnetization in chromia can be electrically switched, this result indicates the possibility of the voltage controlled QAHE at low temperature. We note that a reversible AHE has been realized experimentally at room temperature, although using Pt rather than graphene, as an overlayer on chromia [40].

The asymmetry between the $K$ and $K'$ valleys at the graphene/$Cr_2O_3$ interface gives rise to the valley polarization [41]. The valley polarization appears due to different populations of the two valleys. It can be detected by measuring the longitudinal transport in graphene perpendicular to the AFM domain wall in chromia between two regions with opposite orientation of the Néel vector. Creation and annihilation of the AFM domain wall performs as a valley valve, filtering the valley polarized carriers in graphene. If the chemical potential is engineered to be located within the wider band of the $K$ valley, but below or above the band gap of the $K'$ valley, a perfect valley filtering is expected in the ballistic transport regime. In this case the longitudinal conductance is zero in the presence of the domain wall but non-zero in its absence.

The valley polarization gives rise to the valley Hall effect (VHE) [42]. The VHE can be quantified using a valley Hall conductance (VHC), which is defined as the difference in the AHC between the $K$ and $K'$ valleys, i.e. $\sigma_v = \sigma^K - \sigma^{K'}$. Figure 4c demonstrates that $\sigma_v$ is largest, when the Fermi energy $\varepsilon_f$ lies within only one of the band gaps at either $K$ or $K'$ valley (depending on magnetization orientation). $\sigma_v$ is zero if $\varepsilon_f$ lies within both gaps and is close to a quantized value of $e^2/h$ away from the gaps. The latter originates from the integration of the Berry curvature over the lowest energy band (Figs. 3 a,e), contributing to the VHC $e^2/2h$ from the $K$ valley and $-e^2/2h$ from the $K'$ valley. The VHC of $\sigma_v \approx e^2/h$ represents a spin polarized version of the VHE [42], where the bands contributing to the transport are nearly fully spin-polarized.

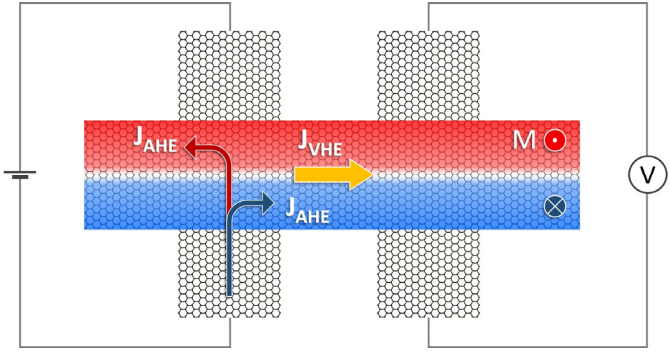

**Figure 5.** Schematic set up for the detection of the valley Hall effect. Shaded regions represent two antiparallel aligned domains of the surface magnetization in chromia. A charge current is generated along the left vertical graphene bar by a current source. It produces an anomalous Hall current, $J_{AHE}$, which flows in the opposite directions above the two AFM domains of $Cr_2O_3$. The AHC has opposite sign in the two domains and thus canceled. On the contrary, a valley Hall current, $J_{VHE}$, is not cancelled and flows in graphene along the domain wall. Such a pure valley current induces a voltage drop between top and bottom terminals of the right vertical graphene bar due to the inverse VHE.

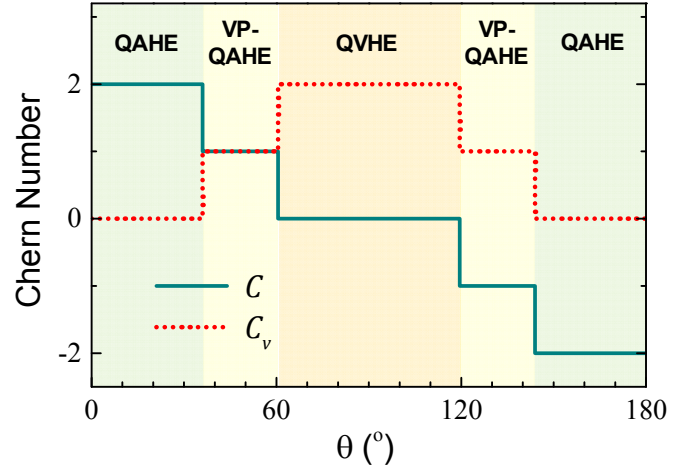

**Figure 6.** Topological phase transformation as a function of magnetization angle $\theta$. Three topological phases are distinguished by the Chern numbers: QAHE ($C = \pm 2, C_v = 0$, indicated by green color); valley-polarized QAHE (VP-QAHE) ($C = \pm 1, C_v = 1$, indicated by yellow color); and QVHE ($C = 0, C_v = 2$, indicated by orange color).

Contrary to $\sigma$, the sign of $\sigma_v$ does not depend on the Néel vector orientation (up or down). Due to this invariance, a pure valley current can be induced in graphene along the domain wall in $Cr_2O_3$. This effect can be realized and measured using a device structure in Figure 5. Here a charge current is generated along the left vertical graphene bar by a current source and produces an anomalous Hall current, $J_{AHE}$, which flows in the opposite directions in the graphene layer regions above the two AFM domains of $Cr_2O_3$. If the domain wall is placed symmetrically (as in Fig. 5), the net anomalous Hall current is cancelled. On the contrary, a valley Hall current, $J_{VHE}$, is not cancelled and flows in graphene along the domain wall. Such a pure valley current induces a voltage drop between top and bottom terminals of the right vertical graphene bar in Figure 5 due to the inverse VHE. This voltage drop can be detected using the non-local measurements, as was reported recently [43-45].

We note that the above scheme assumes that the Fermi level is located in the range of energies where both the AHC and VHC are non-zero. If the Fermi level is engineered to lie above or below the energy gaps, i.e. in the energy region where the AHC vanishes but the VHC does not, the VHC can also be measured using the proposed device scheme without the presence of the domain wall. In this regime, the VHC is expected to be close to $e^2/h$. In fact, these measurements require less stringent conditions compared to those considered above, due to a wider energy range where the VHC is non-zero (as determined by the band splitting at the $K$ and $K'$ points, resulting from the staggered sublattice potential) and the non-local voltage output being independent of the domain structure of $Cr_2O_3$.

### E. Topological phases across a domain wall

The formation of the 180° AFM domain wall also leads to a topological phase transition in graphene, resulting from continues rotation of the surface magnetization. This transition manifests itself in the changing topological invariants across the domain wall. For the surface magnetization pointing normal to the interface, the QAHE is characterized by the Chern number $C = \frac{1}{2\pi}\sum_n \int_{BZ} \Omega^n(\vec{k})d^2k$ (where the summation is performed over all occupied bands) being +2 or –2 for $\theta = 0$ and $\theta = 180°$, respectively. The Chern number is a topological invariant and thus must discontinuously change across the domain wall. At each discontinuity point, the transition must be accompanied by band gap closure.

To observe this transformation, we use Hamiltonian (1) to calculate the evolution of the band structure as a function of magnetization angle $\theta$, assuming for simplicity that the magnetization is uniform. The band structures for different angles $\theta$ are depicted in Supplementary Figures S4 and S4 [35] around the $K'$ and $K$ points, respectively. We find that near the $K'$ point the band gap closes at $\theta \approx 36°$ (Fig. S4b), then it reopens (see Fig. S5c for $\theta \approx 49°$), closes again at $\theta \approx 60°$ (Fig. S4d), and remains open at larger $\theta$ (Figs. S4e-S4i). The band structure around the $K$ point mirrors that near the $K'$ point when the magnetization is flipped from $\theta$ to $180° - \theta$. In this case, the band gap is opened at smaller angles (Figs. S5a-S5e), but it closes at $\theta \approx 120°$ (Fig. S5f), then it reopens (see Fig. S6g for $\theta \approx 131°$) and closes again at $\theta \approx 144°$ (Fig. S5h). Between $36° < \theta < 60°$ and $120° < \theta < 144°$, the maximum band opening is about 0.1 meV at $\theta \approx 49°$ (the $K'$ point) and $\theta \approx 131°$ (the $K$ point). These results indicate that new topological phases emerge in the range of angles $36° \lesssim \theta \lesssim 144°$.

To reveal the nature of these topological phases we calculate the valley dependent Chern numbers, $C_K$ and $C_{K'}$, by integrating the Berry curvature around the $K$ and $K'$ points, respectively. In the calculation, we chose $\varepsilon_f$ to lie within the band gap for $\theta = 0°$, $\theta = 49°$, $\theta = 90°$, $\theta = 131°$, and $\theta = 180°$. We find that the valley dependent Chern numbers are equal for the surface magnetization normal to the interface, i.e. $C_K = C_{K'} = 1$ for $\theta = 0$ and $C_K = C_{K'} = -1$ for $\theta = 180°$. These conditions produce a QAHE phase with the total Chern numbers $C = C_K + C_{K'} = \pm 2$ and the valley Chern number $C_v = C_K - C_{K'} = 0$. However, for the other angles, we obtain that $C_K$ and $C_{K'}$ are different, namely, $C_K = 1$ and $C_{K'} = 0$ at $\theta = 49°$, $C_K = 1$ and $C_{K'} = -1$ at $\theta = 90°$, and $C_K = 0$ and $C_{K'} = -1$ at $\theta = 131°$, signaling for the emergence of new topological phases. The regions around $\theta = 49°$ and $\theta = 131°$ are characterized by $C = \pm 1$, respectively, and $C_v = 1$. This regions exhibit the valley-polarized QAHE phase (VP-QAHE) [46], where the QAHE and the VHE coexist. For magnetization lying in the plane, i.e. $\theta = 90°$, the resulting phase has zero total Chern number, $C = 0$, but a non-zero valley Chern number, $C_v = 2$, indicating the emergence of the quantum valley Hall effect (QVHE) [42]. The corresponding topological phase diagram is depicted in Figure 6. We note that the overall situation is somewhat reminiscent to that predicted for a Bi bilayer where magnetization rotation induced by the spin-orbit torque forces the topological phase transition [36].

## IV. DISCUSSION

The emergence of different topological phases in graphene across the AFM domain wall in chromia is expected to produce chiral edge states (CES) similar to those predicted [8] and observed [47-49] on domain walls of a magnetically doped topological insulator. Two CES are expected to appear along the lines parallel to the domain wall where the topological phase changes. The appearance of the two CES, as well as the QVHE phase, requires a sufficiently wide domain wall with the width larger than the characteristic decay length of the CES into the gapped region, which we estimate to be about 12 nm (Supplementary Information). If the domain wall is not wide enough, the two CES collapse into a single CES and the QVHE phase disappears.

Observation of the predicted phenomena relies on magnetoelectricity of chromia. The latter allows a 180° domain wall of the surface magnetization to be formed by applying voltages of different sign at two local regions to align the Néel vector in opposite directions [50]. The domain wall width can be engineered either by tuning anisotropy with strain or by the split-gate scheme with multiple gates. Furthermore, the locally applied voltages can dynamically control the location of the domain wall [50]. The dynamics may be used to switch paths of injected spin-polarized electrons, which can be detected for readout at specific probes locally attached to graphene along the voltage-controlled domain wall regions in chromia. A further advantage of the proposed scheme is that the interface with the split gates needs neither dopants, adatoms, or external Oersted fields. The magnetoelectric switching is expected to improve the timing of the writing process as to form the domains in memory devices, since it is controlled by applied voltages, rather than by magnetic probe scanning [49]. Specific transport measurements to observe the predicted phenomena can be performed using a four-terminal probe [51-54]. The required $\varepsilon_f$ engineering can be achieved via applied gate voltage [55] or uniaxial pressure [34].

## V. CONCLUSIONS

Our work has outlined a new route for topological antiferromagnetic spintronics. We have predicted the control and manipulation of the topological states in a two-dimensional material via proximity of a magnetoelectric antiferromagnet with the Néel vector being the control parameter. Using the graphene/$Cr_2O_3$ (0001) interface as a model system, we showed the emergence of the unconventional quantum anomalous Hall effect and the spin-polarized valley Hall effect. We predicted the appearance and transformation of the different topological phases in graphene across the 180° AFM domain wall and the emergence of the chiral edge state along the domain wall. These topological properties can be controlled by voltage through magnetoelectric switching of the AFM insulator with no need for spin-orbit torques generated by large currents. Thus, our results provide a viable approach for low-power voltage-controlled topological antiferromagnetic spintronics.

## ACKNOWLEDGMENTS

This work was supported by the National Science Foundation (NSF) through the E2CDA program (grant ECCS-1740136) and the Semiconductor Research Corporation (SRC) through the nCORE program (task 2760.001). AAK and SS acknowledge support from the Department of Energy Early Career Award (grant DE-SC0014189). Computations were performed at the University of Nebraska Holland Computing Center.

## APPENDIX A: DFT COMUTATIONAL DETAILS

DFT calculations are performed using the Vienna *Ab initio* Simulation Package (VASP).[56] We apply local density approximation (LDA) + $U$[57] with $U$ = 4.0 eV and $J$ = 0.58 eV using the projector augmented wave method.[58] In the calculations, we use a $Cr_2O_3$ (0001) slab composed of 8 and 16 atomic layers of O and Cr, respectively (Fig. 1a), assuming the energetically most favorable surface structure terminated with a single Cr layer top and bottom of the slab.[59] A 2×2 graphene sheet, which lattice parameters are strained to the $Cr_2O_3$ (0001) lattice, is placed on top (Fig. 1b) and bottom surfaces of the $Cr_2O_3$ slab. The in-plane placement of graphene on the $Cr_2O_3$ (0001) surface is optimized as described in the Supplementary Material [35]. Periodic boundary conditions along the *z* direction are maintained with a 20Å-thick vacuum buffer layer separating the slabs. Under the constraint of the fixed in-plane lattice constant of $Cr_2O_3$ ($a$ = 4.94 Å), the atomic structures are fully relaxed using the 9×9×1 Γ-centered *k*-point grid and the kinetic energy cutoff of 520 eV. The relaxation is performed until the atomic forces on each atom are converged to better than 0.01 eV/ Å. In the band structure calculations, we increase the *k*-point grid up to 18×18×1.

# Magnetoelectric control of topological phases in graphene

Hiroyuki Takenaka, Shane Sandhoefner, Alexey Kovalev, and Evgeny Y. Tsymbal[*]

*Department of Physics and Astronomy & Nebraska Center for Materials and Nanoscience, University of Nebraska, Lincoln, Nebraska 68588-0299, USA*

## A. Optimization of the graphene/$Cr_2O_3$ (0001) interface structure

The in-plane position of graphene on the $Cr_2O_3$ (0001) surface was optimized by considering three different interface structures: (1) a graphene C atom is atop the Cr atom of the $Cr_2O_3$ surface (Figures 1a,b in the main text), (2) a graphene C atom is atop the O atom in the first O monolayer from the $Cr_2O_3$ surface (Figure S1a), and (3) the Cr surface atom is below the center of a hexagonal ring of graphene (Figure S1b). Starting from these initial configurations, the atomic structure of the whole supercell is relaxed. We find that the lowest total energy structure forms C located atop Cr (Figures 1a,b in the main text). The structure with C atop O atom (Figure S1a) and the structure with Cr under the graphene hollow site, respectively, have the total energy 36 meV and 55 meV higher. In the main text, we focus on the most stable atomic structure to investigate the electric, magnetic, and spin transport of the on the $Cr_2O_3$ (0001) interface.

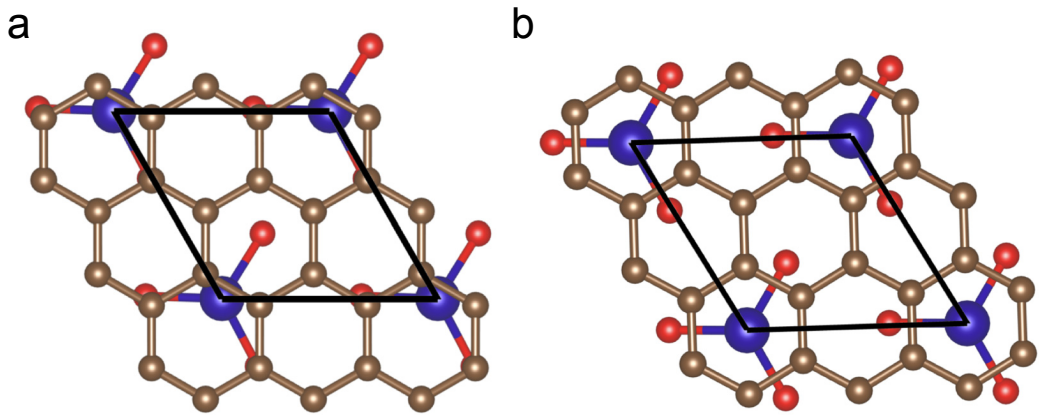

**Figure S1**. Top view of the relaxed atomic structures of the graphene/$Cr_2O_3$ (0001) interface for C atop O subsurface site (**a**) and for Cr under the C hollow site (**b**). Only graphene, Cr surface and O subsurface monolayers are shown. The black solid lines indicate the interface unit cell. Blue, red, and gold spheres indicate Cr, O, and C atoms, respectively.

## B. Band structure of the free-standing graphene layer

Figure S2 shows the DFT-calculated band structure of the free standing graphene layer using the atomic positions of the C atoms in the relaxed graphene/$Cr_2O_3$ (0001) interface structure (Figures 1a,b in the main text). In is seen that the electronic structure projected onto the C $p_z$ orbitals exhibits the Dirac-cone dispersions near the Fermi energy at the $K$ and $K'$ points of the Brillouin zone. For comparison the pristine graphene band structure is overlapped with that for the graphene/$Cr_2O_3$ (0001) interface.

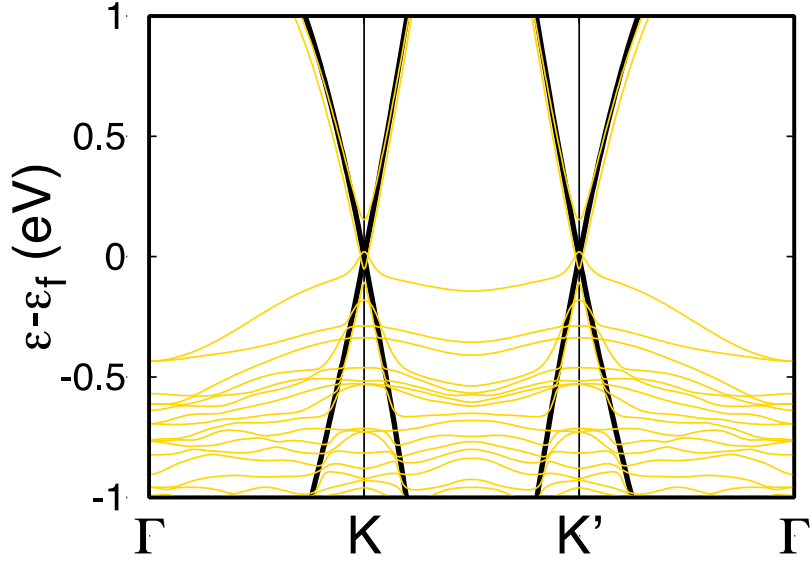

**Figure S2.** The band structure of the free-standing graphene (black thick lines) calculated with DFT using the graphene unit cell as in the graphene/$Cr_2O_3$ (0001) supercell (Figures 1a,b in the main text). Gold thin lines represent the band structure of the interface, the same as in Figures 1c and 1d of the main text.

### C. Fitting of the tight-binding parameters

To obtain the tight-binding parameters of the model Hamiltonian (1) in the main text, we expand the Hamiltonian along the high symmetry *k*-point paths around the *K* and *K'* points,

$$H = -v_f(\eta k_x \sigma_x s_0 + k_y \sigma_y s_0) + \tilde{t}_{so}(\eta \sigma_x s_y - \sigma_y s_x) - \tilde{\lambda}_{nl}(\eta k_x \sigma_x s_z + k_y \sigma_y s_z) + \\ U\sigma_z s_0 + \tilde{\lambda}_{so,A}\eta\sigma_z s_z - \tilde{\lambda}_{so,B}\eta\sigma_0 s_z - J\sigma_0 s_z + \delta J \sigma_z s_z + X s_0 s_0. \quad (S1)$$

Here $\sigma$ and $s$ represent sublattice and spin matrices, respectively; $\eta = 1$ for the *K* point and $\eta = -1$ for the *K'* point. Other parameters in Eq. (S1) are related to those in Eq. (1) of the main text as follows: $t = 2/3\, v_f$, $t_{so} = 2/3\, \tilde{t}_{so}$, $\lambda_{so} = 1/3\sqrt{3}\, \tilde{\lambda}_{so}$, $\lambda_{nl} = 2/3\, \tilde{\lambda}_{nl}$, $J_A = J - \delta J$, $J_B = -(J + \delta J)$. Using this Hamiltonian, we fit the four bands around the Fermi energy which are obtained from our DFT calculation for the graphene/$Cr_2O_3$ (0001) interface. Figure S3 shows the results of the fitting. We note that in the fitting process we focus mostly on the two middle bands, which predominantly contribute to QHC. Although the fitted top and bottom bands somewhat deviate from the DFT results, their contribution to the Berry curvature is relatively small (see Figure S4 below). The model correctly reproduces the reversal of the Néel vector in chromia, which is equivalent to the sign change of the exchange parameters $J$, $\delta J$, and $\tilde{\lambda}_{nl}$ in Hamiltonian (S1). From the fitting, we find the values $v_f = 7.2$ eV, $\tilde{t}_{so} = 1.2$ meV, $\tilde{\lambda}_{nl} = -1.595$ eV, $U = 40.4$ meV, $\tilde{\lambda}_{so,A} = -6.4$ meV, $\tilde{\lambda}_{so,B} = -1.1$ meV, $J = -71.4$ meV, $\delta J = 20$ meV, and $X = 5.6$ meV. We note that this low-energy continuum Hamiltonian fails to capture certain high energy effects.

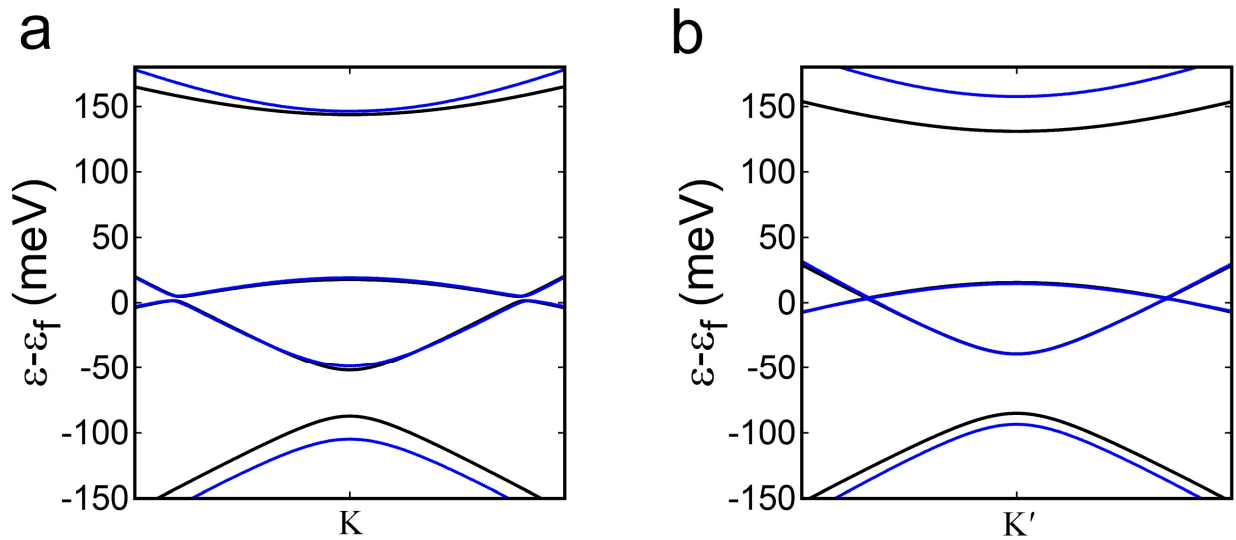

**Figure S3.** Fitting of the DFT band structure of the graphene/Cr$_2$O$_3$ (0001) interface using a model Hamiltonian (S1). The blue lines are the DFT results, and the black lines are the band obtained with the model Hamiltonian. **a**,**b** The band structures around the K (**a**) and K′ (**b**) points.

## D. Band structures as a function of magnetization angle

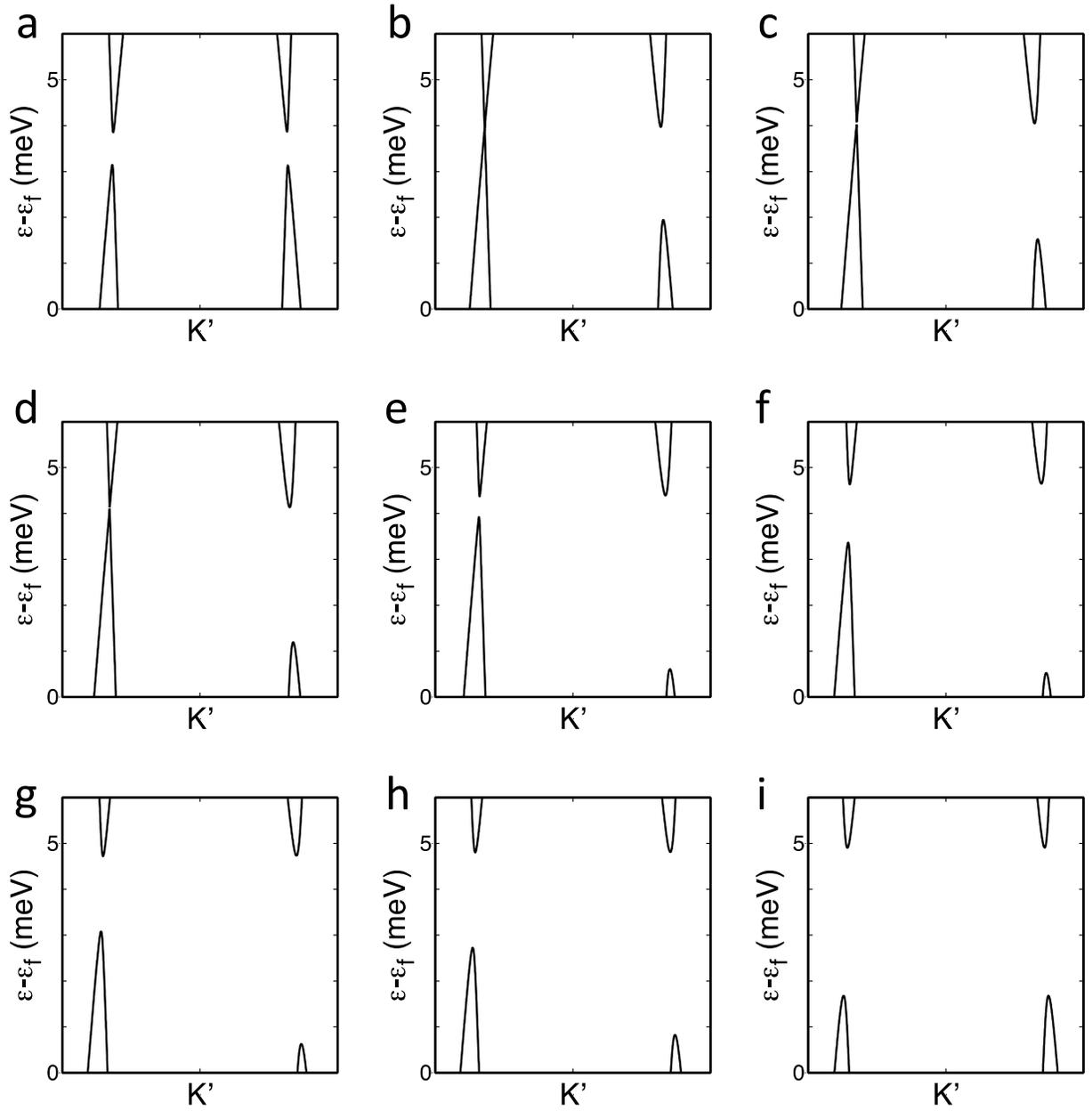

**Figure S4.** The band structures round the $K'$ point at $\theta \approx 0°$ **a**, 36° **b**, 49° **c**, 60° **d**, 90° **e**, 120° **f**, 131° **g**, 144° **h**, and 180° **i**.

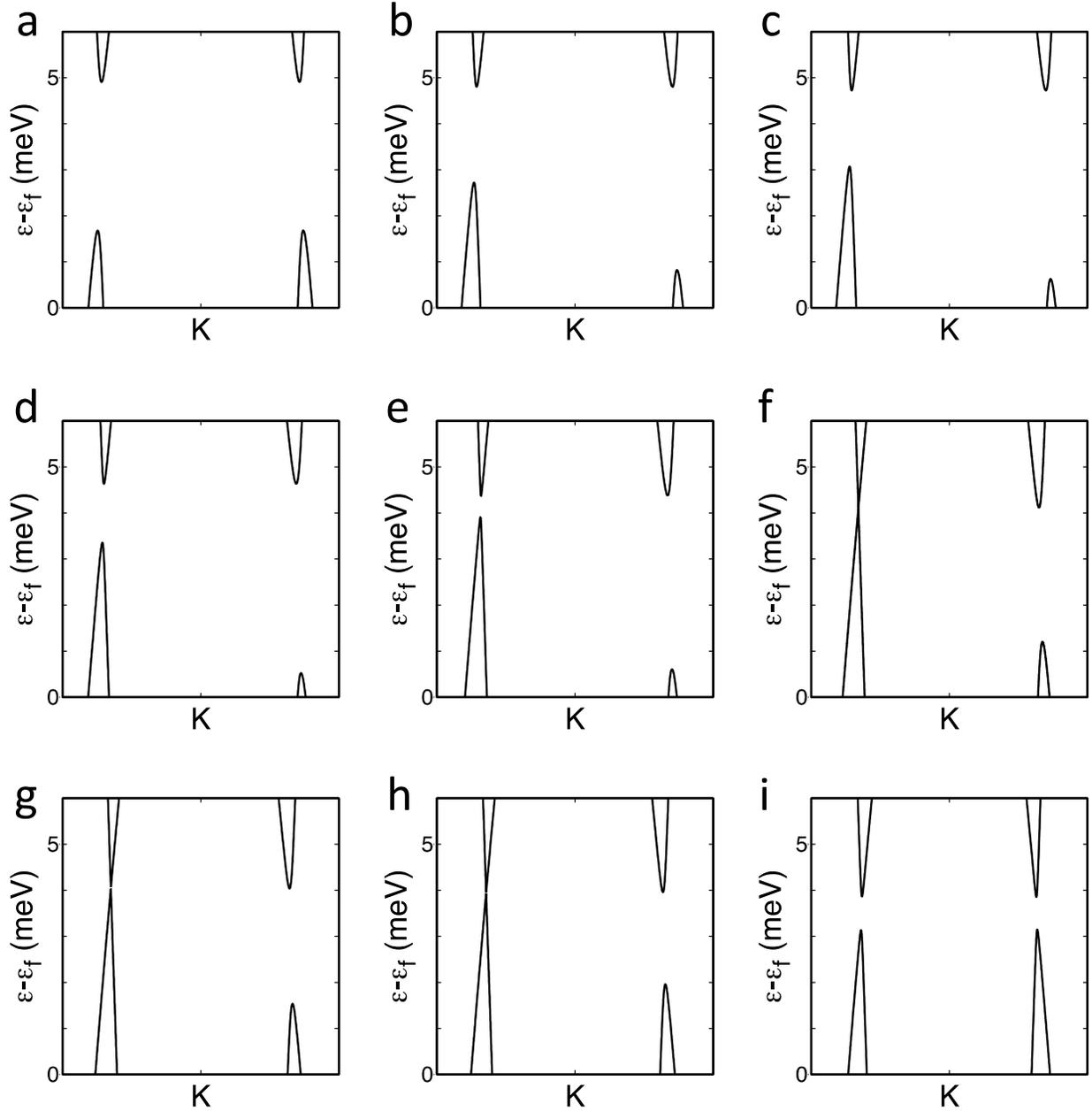

**Figure S5.** The band structures round the $K$ point at $\theta \approx 0°$ **a**, $36°$ **b**, $49°$ **c**, $60°$ **d**, $90°$ **e**, $120°$ **f**, $131°$ **g**, $144°$ **h**, and $180°$ **i**.

### E. CES Decay Length

To estimate the CES decay length (the length over which the probability density decreases to zero), we use a strip geometry with a finite width of 160 sites in the $y$-direction. We apply magnetization in the $+z$-direction and calculate the decay length to be ~12 nm.

5